\documentclass[preprint,aps,12pt,notitlepage,nofootinbib,tightenlines]{revtex4}
\usepackage{amsmath}
\usepackage{enumerate} 
\usepackage{bm}
\usepackage{times}
\usepackage{braket}
\usepackage{color}
\usepackage{epsfig}
\usepackage{slashed}
\usepackage{hyperref}
\usepackage{multirow}
\usepackage{booktabs}
\usepackage{array}
\usepackage{float}
\newcommand{\beq}{\begin{eqnarray}}
\newcommand{\eeq}{\end{eqnarray}}
\newcommand{\be}{\begin{equation}\begin{aligned}}
\newcommand{\ee}{\end{aligned}\end{equation}}

\newcommand{\gev}{\text{GeV}}

\definecolor{Red}{rgb}{1.,0.,0.}

\definecolor{Blue}{rgb}{0.,0.,1.}

\definecolor{nicered}{rgb}{0.7,0.1,0.1}
\definecolor{nicegreen}{rgb}{0.1,0.5,0.1}
\def\lsim{ {\ \lower-1.2pt\vbox{\hbox{\rlap{$<$}\lower6pt\vbox{\hbox{$\sim$}}}}\ } }
\def\gsim{ {\ \lower-1.2pt\vbox{\hbox{\rlap{$>$}\lower6pt\vbox{\hbox{$\sim$}}}}\ } }

\hypersetup{colorlinks,citecolor=nicegreen,linkcolor=nicered}
\begin{document}
\title{Search for pair production of the heavy vectorlike top partner in same-sign dilepton signature at the HL-LHC}
\author{Xiao-Min Cui, Yu-Qi Li, Yao-Bei Liu\footnote{E-mail: liuyaobei@hist.edu.cn}}
\affiliation{Henan Institute of Science and Technology, Xinxiang 453003, People's Republic of China}%%

%\date{\today}
\begin{abstract}
New vectorlike quarks are predicted in many new physics scenarios beyond the Standard Model~(SM)  and could potentially be
discovered at the LHC.  Based on a simplified model including a singlet vectorlike top partner with charge $2/3$, we investigate the process $pp \to TT$ via a $t$-channel induced by the couplings between the top partner with the first-generation SM quarks. We calculate   the production cross section and further study the observability of the heavy top partner in the channel $T\to Wq$ at the high-luminosity LHC (HL-LHC) using final states with same-sign dileptons (electrons or muons), two jets, and missing transverse momentum.
 At the 14 TeV
LHC with an integrated luminosity of 3000 fb$^{-1}$,  the $2\sigma$ exclusion limits, as well as the $5\sigma$ discovery reach in the parameter plane of the two variables  $g^{\ast}-R_L$, are respectively obtained at the HL-LHC.  We also obtain the $2\sigma$ exclusion limit on the coupling strength parameter  $g^{\ast}$ in the case in which the vectorlike top partner is coupled only to the first-generation quarks.
\end{abstract}

\maketitle

%%====================================================================
\newpage
\section{Introduction}
Although the Standard Model (SM) has proved itself with great success, a theory beyond the SM (BSM) is necessary from both the theoretical and experimental points of view, one of which is the so-called  gauge hierarchy problem~\cite{DeSimone:2012fs}. Many new physics models BSM, such as little Higgs~\cite{ArkaniHamed:2002qy,Han:2003wu,Chang:2003vs}, composite Higgs~\cite{Agashe:2004rs}, and other extended models~\cite{He:1999vp,Wang:2013jwa,He:2001fz,He:2014ora}, have been proposed to solve this problem by introducing a spontaneously broken global symmetry, leading the Higgs boson to be a
pseudo Goldstone boson. New vectorlike top partners~(VLQ-$T$) are generally predicted in these BSM models, which are color-triplet fermions but with its left- and right-handed
components transforming in the same way under the gauge group $SU(2)\times U(1)$~\cite{Buchkremer:2013bha,Aguilar-Saavedra:2013qpa}.
 A common feature is that they are assumed to decay into a SM quark and a gauge boson or
Higgs boson, which can  generate characteristic signatures at hadron colliders~(see, for example~\cite{Cacciapaglia:2011fx,Okada:2012gy,Backovic:2014uma,Barducci:2017xtw,Cacciapaglia:2018qep,
Liu:2017rjw,Liu:2017sdg,Liu:2019jgp,Tian:2021oey,Tian:2021nmj,Yang:2021btv,Moretti:2016gkr,Moretti:2017qby,
Carvalho:2018jkq,Benbrik:2019zdp,Aguilar-Saavedra:2019ghg,Buckley:2020wzk,Brown:2020uwk,Deandrea:2021vje,Belyaev:2021zgq,Dasgupta:2021fzw,Han:2022npb,Cacciapaglia:2021uqh,Bhardwaj:2022wfz,Han:2022jcp,Verma:2022nyd}).

 From the experimental point of view, vectorlike quarks~(VLQs) are still allowed by present searches, unlike the fourth
generation of quarks with chiral couplings, which is ruled out by electroweak precision measurements~\cite{Kribs:2007nz,Banerjee:2013hxa,Cao:2022mif}, and
by the measured properties of the SM Higgs boson~\cite{ATLAS:2016neq,Eberhardt:2012sb,CMS:2013zma}. VLQs can evade such exclusion bounds because they are not chiral,
a priori, and do not have to acquire their mass via the Higgs mechanism. Therefore, such new particles are receiving a lot of
attention at the LHC. Up to now, searches at the LHC for VLQ-$T$ have been performed and presented by
the ATLAS and CMS Collaborations, with  the lower mass bounds on $T$ reaching up to about
$740-1370$~GeV at 95\% confidence level (C.L.), depending on the $SU(2)$ multiplets they belong to and different decay modes ~\cite{Aaboud:2018pii,CMS:2019eqb}. Besides, such VLQ-$T$ can also be singly produced at the LHC via  their electroweak~(EW) coupling with SM quarks and weak bosons, which depends on the
strength of the interaction between the VLQ-$T$  and the weak gauge bosons. Current
searches for single production of VLQ-$T$ have placed limits on the production
cross sections for their masses between 1 and 2~TeV at 95\% C.L. for various EW coupling parameters~\cite{CMS:2017gsh,CMS:2017voh,CMS:2019afi,ATLAS:2022ozf}.

Typically, most of the phenomenological studies are based on the assumption that the VLQ-$T$ only couple to the  third-generation quarks, since this is the scenario least constrained by previous measurements~\cite{Buchkremer:2013bha}.
Considering the constraints from flavor physics~\cite{Cacciapaglia:2010vn,Botella:2012ju,Cacciapaglia:2015ixa,Alok:2015iha,Ishiwata:2015cga,Botella:2016ibj,Vatsyayan:2020jan,Branco:2021vhs,Accomando:2022ouo,Balaji:2021lpr},
the VLQ-$T$ can mix in a sizable way with
lighter quarks, which could have a
severe impact on electroweak vectorlike quark processes
at the LHC~\cite{Atre:2011ae,Basso:2014apa,Liu:2016jho} and the Large Hadron Electron Collider~\cite{Han:2017cvu,Zhang:2017nsn,Gong:2020ouh}.
This is particularly of
interest for couplings to first-generation quarks, where amplitudes involving VLQ-$T$ couplings
direct to initial-state up quark become significant due to large high-$x$ valence-quark densities.
The future high-luminosity LHC (HL-LHC) is expected to reach 3000~fb $^{-1}$~\cite{Apollinari:2015wtw}, which will be very beneficial for discovering possible new physical signals even for small production rates. Recently, Zhou and Liu~\cite{Zhou:2020ovl,Zhou:2020byj} studied a new decay channel of the top partner mediated by the heavy Majorana neutrino~($T\to b\ell^{+}\ell^{+}jj$),  which can be used to
probe the top partner and test the seesaw mechanism simultaneously at the HL-LHC by searching for final same-sign dileptons.
 In this work, we study the pair production of the VLQ-$T$ at the HL-LHC in a model-independent way through the process $pp \to TT$ with the decay channel $T\to W^{+}q(\to \ell^{+} \nu_{\ell}q)$,  which induced the final states with two leptons of the same
electric charge (electrons or muons), two jets, and missing transverse momentum.

The paper is arranged as follows. In Sec. II, we briefly review the simplified model including the singlet VLQ-$T$  and calculate its pair production involving the mixing with both the first- and third-generate quarks. In Sec.~III, we discuss the observability of the VLQ-$T$ through the process $pp \to TT\to \ell^{+}\ell^{+}jj+\slashed E_T$
at the HL-LHC. Finally, conclusions are presented in Sec.~IV.

\section{Top partner in the simplified model}
\subsection{An effective Lagrangian for singlet VLQ-$T$}
Buchkremer {\it et al.}~\cite{Buchkremer:2013bha} proposed a generic parametrization of an effective Lagrangian for vectorlike quarks with different electromagnetic charge, where they considered vectorlike quarks embedded in general
representations of the weak $SU(2)$ group. In particular, vectorlike quarks which can mix and decay
directly into SM quarks of all generations are included. Particularly interesting for our
purposes is the case in which the VLQ-$T$ is an $SU(2)$ singlet and can mix and decay directly into the first and third generation of SM quarks.
The Lagrangian parametrizes the VLQ-$T$ couplings to quarks and electroweak boson can be expressed as\footnote{Note that the model file of the singlet VLQ-$T$ is publicly available online in the FeynRules repository~\cite{http}.}
\begin{eqnarray}
{\cal L}_{T} =&& \frac{g^{\ast}}{2}\left\{ \sqrt{\frac{R_L}{1+R_L}}\frac{g}{\sqrt{2}} \right.[\bar{T}_{L}W_{\mu}^{+}
    \gamma^{\mu} d_{L}]  + \sqrt{\frac{1}{1+R_L}}\frac{g}{\sqrt{2}} [\bar{T}_{L} W_{\mu}^{+} \gamma^{\mu} b_{L}]\nonumber \\
  && + \sqrt{\frac{R_L}{1+R_L}}\frac{g}{2\cos \theta_W} [\bar{T}_{L} Z_{\mu}^{+} \gamma^{\mu} u_{L}] +\sqrt{\frac{1}{1+R_L}}\frac{g}{2\cos \theta_W} [\bar{T}_{L} Z_{\mu}^{+} \gamma^{\mu} t_{L}] \nonumber \\
  &&  \left. - \sqrt{\frac{R_L}{1+R_L}}\frac{M_{T}}{v}[\bar{T}_{R}Hu_{L}] - \sqrt{\frac{1}{1+R_L}}\frac{M_{T}}{\upsilon}[\bar{T}_{R}Ht_{L}]- \sqrt{\frac{1}{1+R_L}}\frac{m_{t}}{\upsilon} [\bar{T}_{L}Ht_{R}] \right\}+ H.c. ,
  \label{TsingletVL}
\end{eqnarray}
where
 $g$ is the $SU(2)_L$ gauge coupling constant, $\theta_W$ is the Weinberg angle, and $\upsilon \simeq 246$~GeV. Besides the VLQ-$T$ mass $M_T$, there are the following two free parameters:
\begin{itemize}
\item
 $g^{\ast}$, the coupling
strength to SM quarks in units of standard couplings, which is relevant only to the EW couplings.
\item
$R_L$, the generation mixing coupling parameter, which controls the
share of the VLQ-$T$ coupling between first- and third-generation quarks. In the extreme case, $R_L = 0$ and $R_L = \infty$, respectively, correspond to
coupling to third-generation quarks and the first generation of quarks only.
\end{itemize}
%%Fig.1 %%%%%%%%%%%%%%%%%%%%
\begin{figure}[thb]
\begin{center}
\vspace{-0.5cm}
\centerline{\epsfxsize=10cm \epsffile{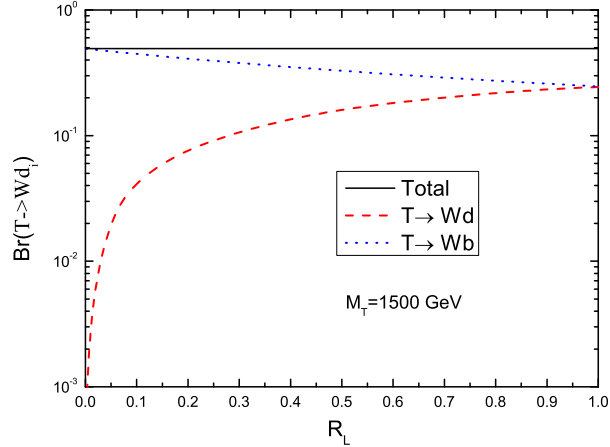}}
\caption{Branching ratios of the decay mode $T\to Wd_{i}$ as a function of  the mixing parameter $R_L$ for $M_T=1500$~GeV.}
\label{fig:br}
\end{center}
\end{figure}

According to the above discussions, VLQ-$T$ has three typical decay modes: $Wd^{}_i$, $Zu^{}_i$, and $Hu^{}_i$, where $i = 1, 3$ is the
index for the first and third generations of the SM fermions.
In the limit of $M^{}_T \gg m^{}_t$, the partial widths can be approximately written as
\begin{eqnarray}
 \Gamma ( T \to Wd^{}_i) &\simeq& \frac{c_{i} e^2 (g^{\ast})^{2} M_{T}^{3} } {256 \pi \sin^2\theta_W m_W^2 } \;,
 \\
 \Gamma ( T \to Zu^{}_i ) &\simeq& \frac{c_{i} e^2 (g^{\ast})^{2} M_{T}^{3} } {512 \pi \sin^2\theta_W m_W^2 } \;,
 \\
  \Gamma ( T \to Hu^{}_i ) &\simeq& \frac{c_{i} e^2 (g^{\ast})^{2} M_{T}^{3} } {512 \pi \sin^2\theta_W m_W^2 } \;.
 \end{eqnarray}
 where $c^{}_i = 1/\left(1+R^{}_L\right)$ for $t$ and $b$ quarks, and $c^{}_i =  R^{}_L/\left(1+R^{}_L\right)$ for $u$ and $d$ quarks,
From the above equations, we can see that the branching fractions of $T$ into $Hu_{i}$, $Zu_{i}$ and $Wd_{i}$ reach a good
approximation for a large mass of VLQ-$T$, given by the ratios $1 : 1 : 2$ as expected from the Goldstone boson equivalence theorem~\cite{He:1992nga,He:1993yd,He:1994br,He:1996rb,He:1996cm}.  A full study of the precision bounds
of this particular model is beyond the scope of this paper, as we use this model only as illustration for VLQ-$T$ search strategies. These parameters can be constrained by the flavor physics and the oblique parameters. Here we consider a phenomenologically
guided limit $g^{\ast}\leq 0.5$ and $0\leq R_L\leq 1$. We also consider the case of $R_L = \infty$ in later discussions.

The branching ratios of the decay mode $T\to Wd_{i}$ are plotted as functions of the mixing parameter $R_L$ in Fig.~\ref{fig:br}.
For $M_T = 1500$ GeV, we can obtain that the branching ratio of $Br \left(T\rightarrow Wd_i\right)$ is approximate equal to 50\%.
As expected, the branching ratios of the first-generation quark vanish rapidly when $R^{}_L$ approaches zero. For $R_L=1$, the branching ratios that decay into the first- and third-generation quarks are approximately equal.
Hence,  we choose the $Wd_{i}$ channel to study the possibility of detecting the signals of VLQ-$T$ at the LHC in our work.

\subsection{Pair production of  VLQ-$T$ at the LHC}
\begin{figure}[htb]
\begin{center}
\vspace{-1.0cm}
\centerline{\epsfxsize=14cm \epsffile{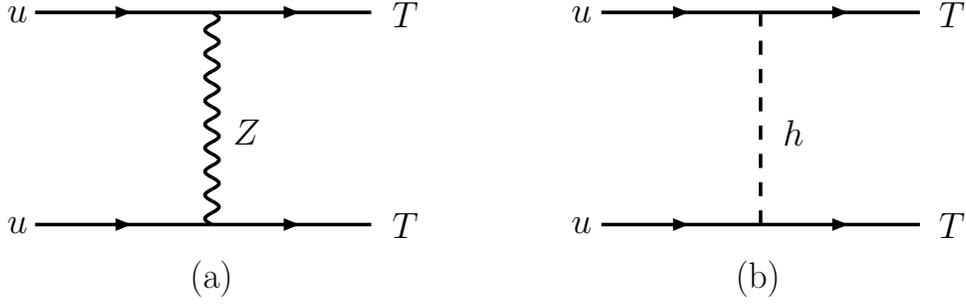}}
\vspace{-14.5cm}
\caption{Feynman diagrams for the process $uu \to TT$ at the
LHC.}
\label{fey}
\end{center}
\end{figure}

Owing to the interaction with the first-generation quarks, the top partner can be pair produced by $t$-channel exchange of the $Z$ gauge boson and Higgs boson. The relevant Feynman diagrams are presented in Fig.~\ref{fey}.
%%%%%%%%%%%%%%%%%%%%%%%%%%%%%%%%%%%%%%%%%%%%%%%%%%%%
%%%%%%%%%%%%%%%%%%%%%%%%%

%%%%%%%%%%%%%%%%%%%%%%%%%

\begin{figure}[thb]
\begin{center}
\vspace{-0.5cm}
\centerline{\epsfxsize=9cm \epsffile{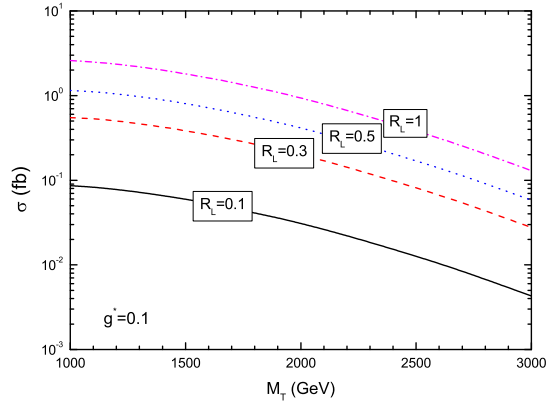}}
\caption{Cross sections of the process $pp\to TT$ as functions of $M_T$ for $g^{\ast}=0.1$ and different values of $R_L$ at the 14~TeV LHC.  }
\label{cross}
\end{center}
\end{figure}

The production cross section $ \sigma(pp\rightarrow TT)$ is plotted in Fig.~\ref{cross}, as a function of the  mass $M_T$ for for $g^{\ast}=0.1$ and several values of $R_L$ at the 14 TeV LHC.
The leading-order~(LO) cross sections are obtained using MadGraph5-aMC$@$NLO~\cite{Alwall:2014hca} with NNPDF23L01 parton distribution functions (PDFs)~\cite{Ball:2014uwa} taking the default renormalization and factorization scales.
It is clear that the values of the cross sections are very sensitive to $R_L$. This implies that the mixing with the first generation can largely enhance the pair production due to the large quark PDFs. Besides, the cross section falls slowly for a higher mass.
Certainly, for the fixed VLQ-$T$ mass,  the production cross section is proportional to the values of $(g^{\ast})^{4}$. Thus, the above advantages make it an ideal process for discovery of heavy VLQ-$T$ with small coupling to the first-generation quarks.
%%Fig.3 %%%%%%%%%%%%%%%%%%%%
%%%%%%%%%%%%%%%%%%%%%%%%%
\section{Event generation and discovery potentiality}

Next, we perform the
Monte Carlo simulation and explore the sensitivity of the VLQ-$T$ at the 14~TeV LHC through the channel,
\begin{equation}\label{signal}
pp \to T(\to Wd_{i})T(\to Wd_{i})\to \ell^{+}\ell^{+}jj+\slashed E_T,
\end{equation}
where $\ell= e, \mu$.

 For the above same-sign dilepton  final states, the major SM backgrounds at the LHC come from prompt multileptons (mainly from
events with $t\bar{t}W^{+}$ and $W^{+}W^{+}+$jets) and nonprompt leptons (mainly from events with jets of heavy flavor, such as $t\bar{t}$).
Other processes, such as the  $t\bar{t}Z$, triboson events, $ZZjj$, and $W^\pm+$jets are not included in the analysis owing to the negligible cross sections resulting from application of the cuts. To be exact, opposite-sign dileptons, one of which is mismeasured, should also constitute our backgrounds but, as the rate of mismeasurement for muons, is
generally low enough that we ignore its effects.
The QCD next-to-leading-order~(NLO) prediction for pair production is calculated in Ref.~\cite{Fuks:2016ftf}. Here we take the conservative value of the $K$-factor as 1.3 for the signal.
To account for contributions from higher-order QCD corrections,
 the cross sections of dominant backgrounds at LO are adjusted to NLO by means of
$K$ factors, which are 1.04 for $W^{+}W^{+}jj$~\cite{Jager:2009xx,Melia:2010bm} and 1.22 for $t\bar{t}W^{+}$~\cite{Campbell:2012dh}. The dominant  $t\bar{t}$ background is normalized to the NNLO QCD cross section of
953.6 pb~\cite{Czakon:2013goa}. It should be noted that we assume that the kinematic distributions are
only mildly affected by these higher-order QCD effects. Therefore, for
simplicity, we rescale the above distributions by using constant
bin-independent $K$ factors.

Signal and background events are generated at LO  using
MadGraph5-aMC$@$NLO. As a reference point, we set
a benchmark value of $g^{\ast}=0.1$ and $R_L=1$. Analogously, our benchmark
points  in the mass axis read $M_T=$1500 and 2000~GeV.  However, we will present the reach later in the $g^{\ast}-R_L$
plane. Then we pass the parton-level events to Pythia 8.20 \cite{pythia8} and Delphes 3.4.2~\cite{deFavereau:2013fsa} for performing the parton
shower and fast detector
simulations, respectively. The anti-$k_{t}$ algorithm
\cite{Cacciari:2008gp} with parameter $\Delta R=0.4$ is used to reconstruct jets. Finally, event analysis is performed by using MadAnalysis5~\cite{ma5}.

To identify objects, we choose the basic cuts at parton level for the signals and SM backgrounds as follows:
 \be
p_{T}^{\ell/j}>~50~\gev,\quad
 |\eta_{\ell/j}|<~2.5, \quad
 \Delta R_{ij} > 0.4,\\
  \ee
where $\Delta R=\sqrt{\Delta\Phi^{2}+\Delta\eta^{2}}$ is the separation in the rapidity-azimuth plane and $p_{T}^{\ell/ j}$ and $|\eta_{\ell/j}|$ are the transverse momentum and pseudorapidity of the leptons and jets, respectively.
%%%%%%%%%%%%%%%%%%%%%%%%%%%%%%%%%%%%%%%%%%%%%%%%%%
\begin{figure*}[htb]
\begin{center}
\centerline{\hspace{2.0cm}\epsfxsize=9cm\epsffile{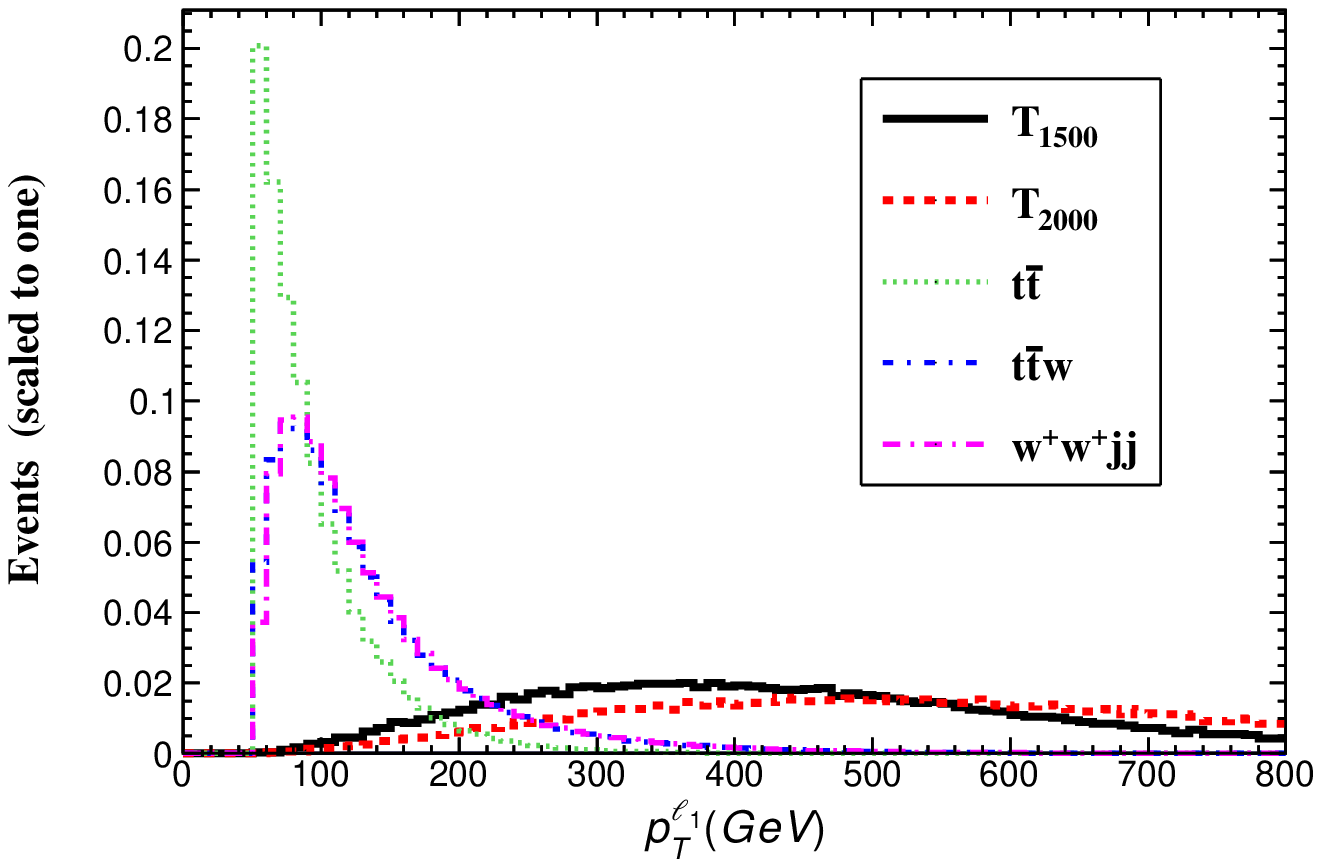}
\hspace{-2.0cm}\epsfxsize=9cm\epsffile{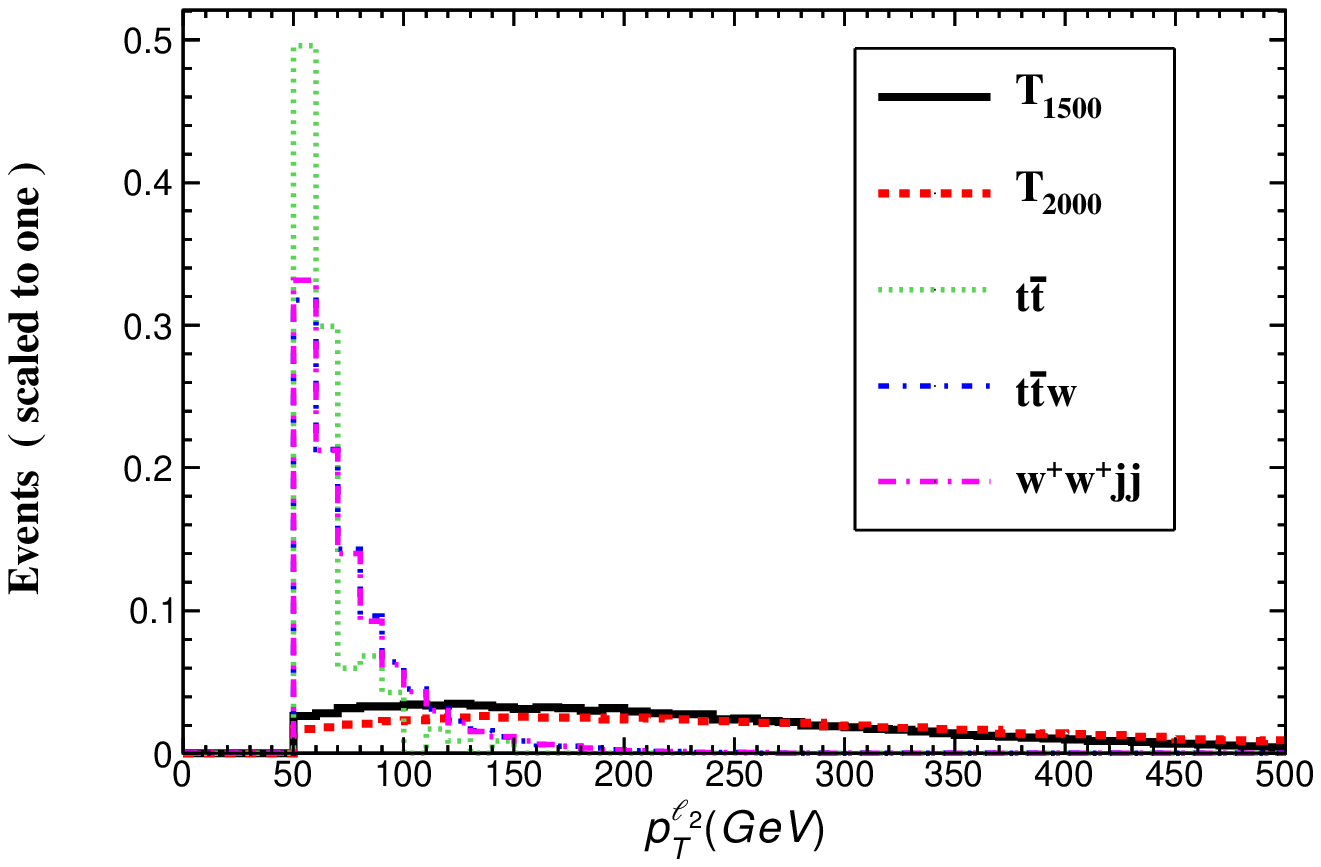}}
\centerline{\hspace{2.0cm}\epsfxsize=9cm\epsffile{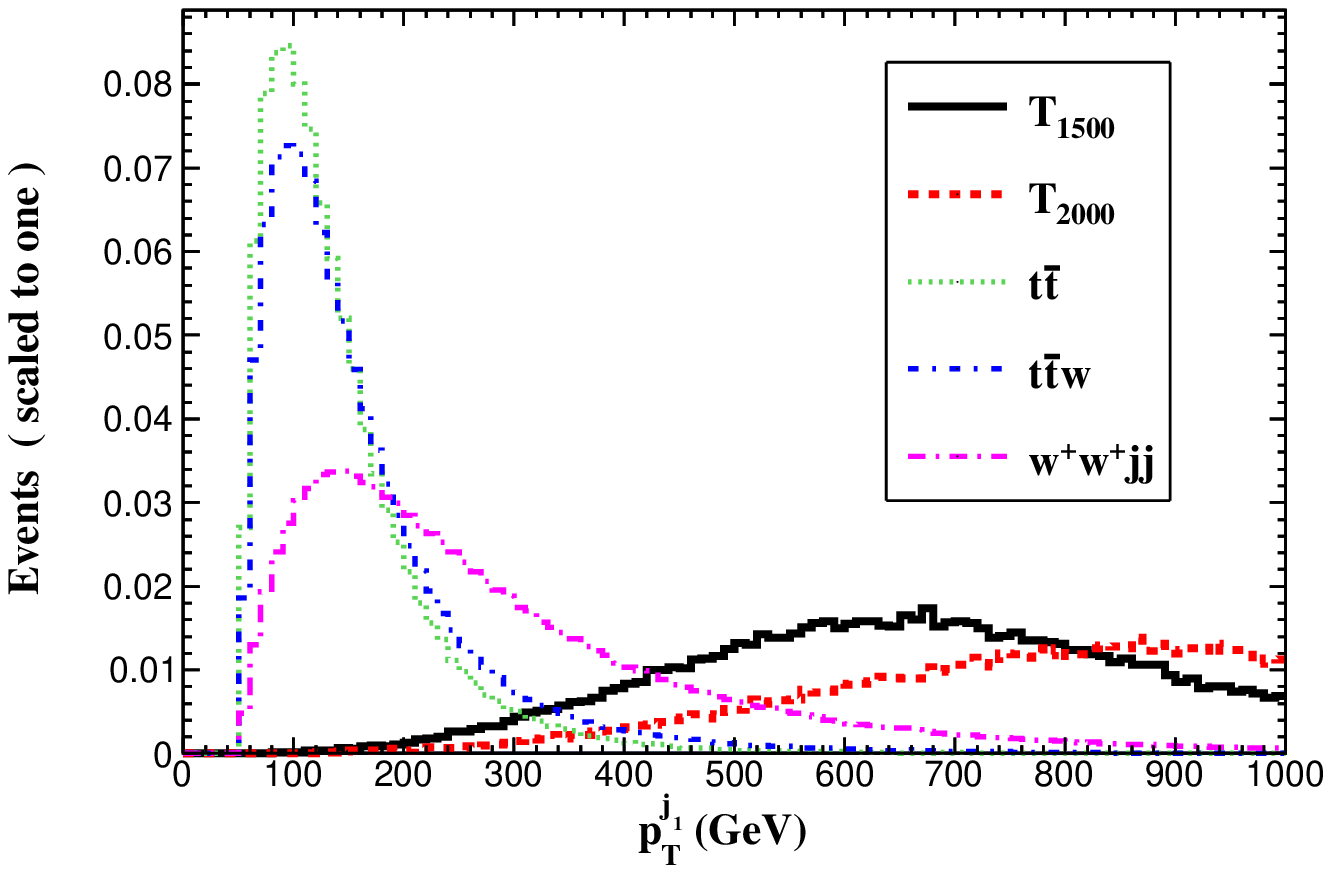}
\hspace{-2.0cm}\epsfxsize=9cm\epsffile{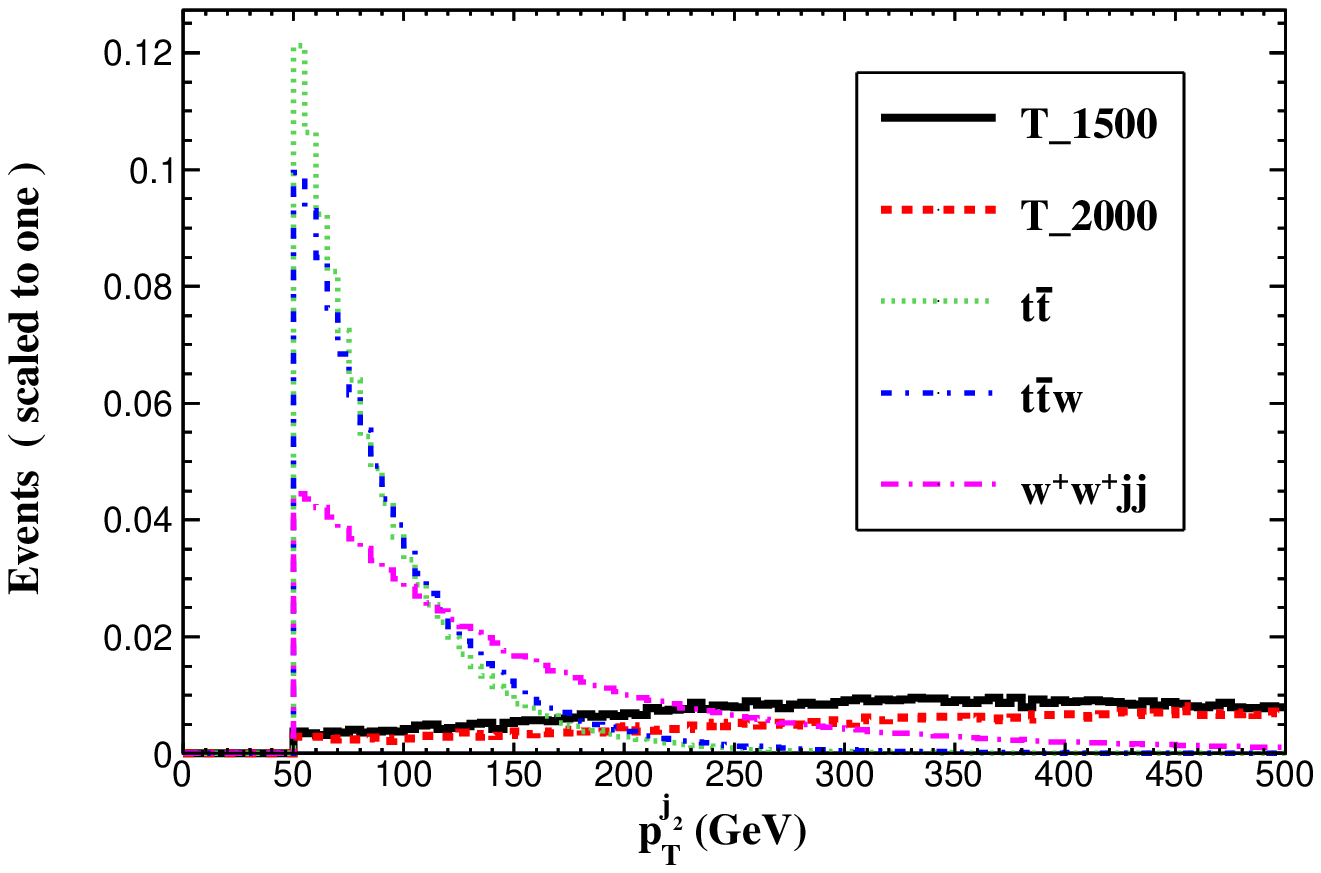}}
\centerline{\hspace{2.0cm}\epsfxsize=9cm\epsffile{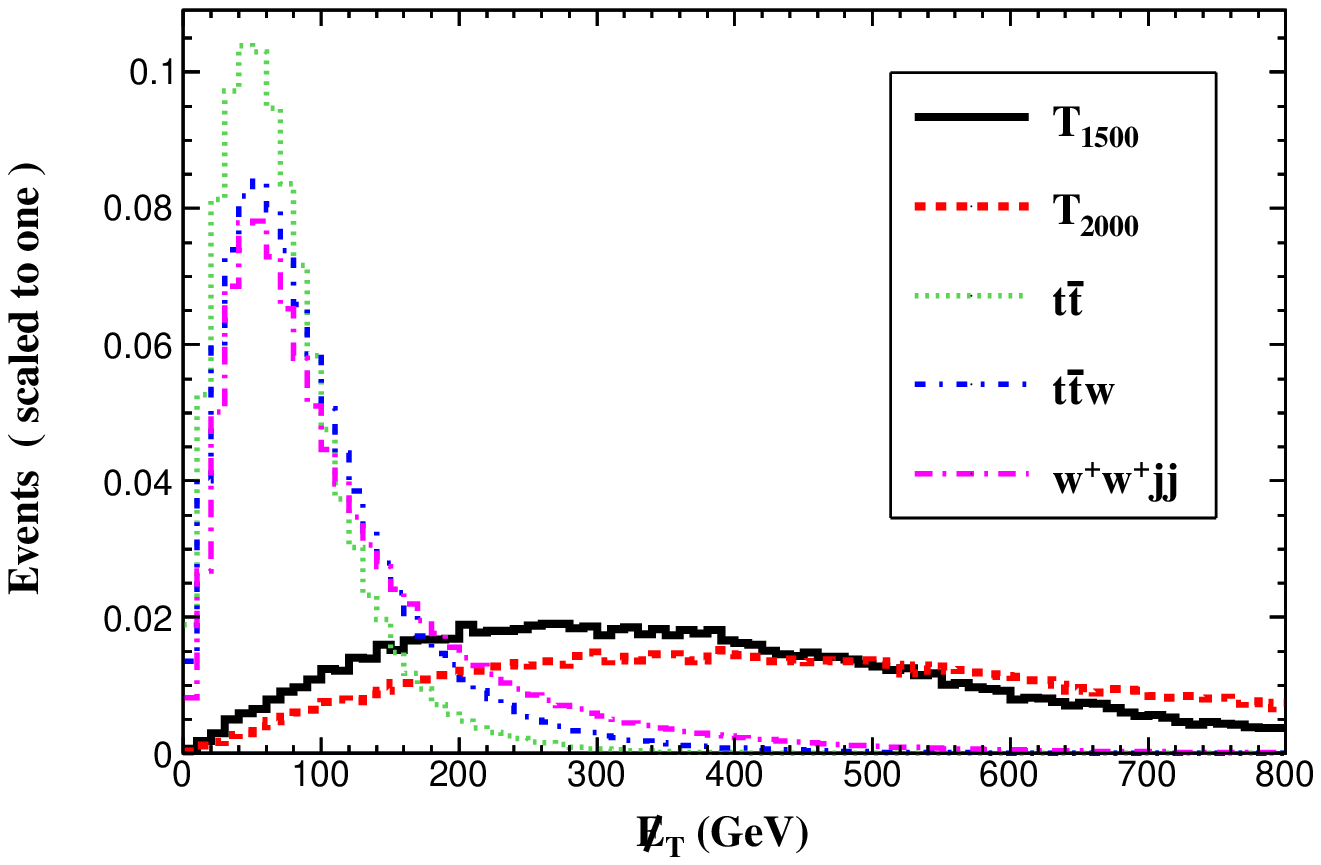}
\hspace{-2.0cm}\epsfxsize=9cm\epsffile{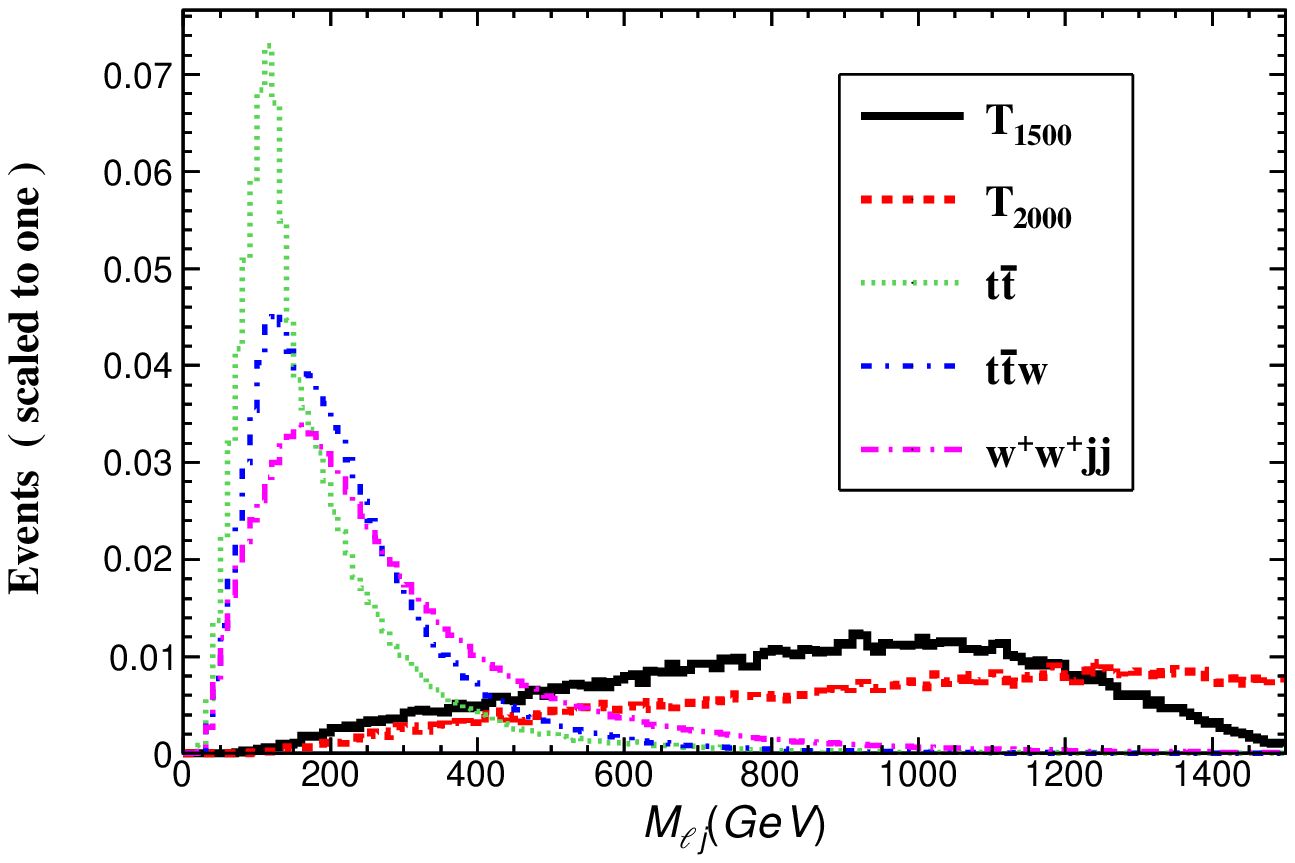}}
\caption{Normalized distributions for the signals (with $m_{T}=1500$ and 2000 GeV) and SM backgrounds. }
\label{fig4}
\end{center}
\end{figure*}

Owing to the larger mass of VLQ-$T$, the decay products are highly boosted. Therefore,
the $p_{T}^{l/j}$ peaks of the signals are larger than those of the SM backgrounds.
In Fig.~\ref{fig4}, we plot some differential distributions  for  signals and SM backgrounds at the LHC, such as the transverse momentum distributions of the leading and subleading leptons~($p_{T}^{\ell_{1,2}}$),  the transverse momentum distributions of the leading and subleading jets~($p_{T}^{j_{1,2}}$), the missing transverse energy $\slashed E_{T}$, and the invariant mass distribution for the final $j\ell$ system $M_{\ell j}$.
Based on these kinematical distributions, we apply
the following kinematic cuts to the events to distinguish the signal from the SM backgrounds.
 \begin{enumerate}[(a)]
\item Cut 1: There are exactly two same-sign isolated leptons~$[N(\ell^{+})=2]$  and at least two jets~$[N(j)\geq2]$
\item Cut 2: The transverse momenta of the leading and subleading leptons and jets are required $p_{T}^{l_{1,2}}> 200~(100) \rm ~GeV$ and  $p_{T}^{j_{1,2}}> 300~(150) \rm ~GeV$. Besides, the invariant mass of two jets are required $M_{jj}> 200 \rm ~GeV$ to reduce the background from $W$-boson decays.
\item Cut 3: The transverse missing energy is required  $\slashed E_{T}> 200 \rm ~GeV$.
\item Cut 4: The invariant mass of final system $M_{\ell j}$ is required to have $M_{\ell j}> 600 \rm ~GeV$.
\end{enumerate}

\begin{table}[htb]
\centering %
\caption{Cut flow of the cross sections (in fb) for the signals and SM backgrounds at the 14 TeV LHC and two typical VLQ-$T$ quark masses. Here we take the parameters $g^{\ast}=0.1$ and $R_L=1.0$. \label{cutflow}}
\vspace{0.8cm}
\begin{tabular}{p{1.6cm}<{\centering} p{2.0cm}<{\centering} p{2.0cm}<{\centering}p{0.3cm}<{\centering} p{2.5cm}<{\centering}  p{2.5cm}<{\centering} p{2.5cm}<{\centering} }
\toprule[1.5pt]
 \multirow{2}{*}{Cuts}& \multicolumn{2}{c}{Signals}&\multicolumn{4}{c}{Backgrounds}  \\ \cline{2-3}  \cline{5-7}
&1500 GeV &2000 GeV&& $t\bar{t}$  &$t\bar{t}W^{+}$  &$W^{+}W^{+}jj$\\    \cline{1-7} \midrule[1pt]
Basic&0.014&0.0069&&1221&1.54&0.43\\
Cut 1&0.014&0.0069&&1.06&1.29&0.43\\
Cut 2&0.0095&0.0056&&$8.1\times 10^{-4}$&0.007&0.013\\
Cut 3 &0.0074&0.0049&&$2.4\times 10^{-4}$&0.002&0.0049\\
Cut 4 &0.0056&0.0041&&$4.6\times 10^{-5}$&$3.6\times10^{-4}$&0.0014\\
\hline
Efficiency &41\%&59\%&&$3.8\times 10^{-8}$&0.023\%&0.33\% \\
\bottomrule[1.5pt]
\end{tabular}
 \end{table}

We present the cross sections of three typical signal ($M_T=1500, 2000$~GeV) and the relevant
backgrounds after imposing
the cuts in Table~\ref{cutflow}.  Among the three kinds of SM backgrounds,
we can see from Table~\ref{cutflow} that the dominant one is the $t\bar{t}$ events with the basic cut. The first two cuts on
numbers of final same-sign leptons and transverse momenta of leptons and jets can greatly suppress the $t\bar{t}$ events, and other SM backgrounds to the same
order as the signal remain. Then the large $\slashed E_{T}$ requirement can cut about 70\% SM backgrounds
while keeping 80\% signal events. All backgrounds are suppressed very efficiently at the end of the cut flow, while the signals still have a relatively good efficiency. The dominant SM  background comes from the $W^{+}W^{+}jj$ process, with a cross section of $1.4\times10^{-3}$~fb.

It should be noted that we have not considered the pileup effects, which
is important for a fully realistic simulation and needs appropriate  removal techniques~\cite{Cacciari:2007fd,Krohn:2013lba,Berta:2014eza}.
 However, we expect that such effects can be limited on our results since the event selection is based
on two same-sign hard leptons.

The median expected significance for discovery and exclusion can be approximated by~\cite{Cowan:2010js}
\be
\mathcal{Z}_\text{disc} &=
  \sqrt{2\left[(s+b)\ln\left(\frac{(s+b)(1+\delta^2 b)}{b+\delta^2 b(s+b)}\right) -
  \frac{1}{\delta^2 }\ln\left(1+\delta^2\frac{s}{1+\delta^2 b}\right)\right]} \\
   \mathcal{Z}_\text{excl} &=\sqrt{2\left[s-b\ln\left(\frac{b+s+x}{2b}\right)
  - \frac{1}{\delta^2 }\ln\left(\frac{b-s+x}{2b}\right)\right] -
  \left(b+s-x\right)\left(1+\frac{1}{\delta^2 b}\right)},
 \ee
with
 \be
 x=\sqrt{(s+b)^2- 4 \delta^2 s b^2/(1+\delta^2 b)}.
  \ee
  In the idealized limit of a perfectly known background prediction, $\delta=0$, these expressions would reduce to
\be
 \mathcal{Z}_\text{disc} &= \sqrt{2[(s+b)\ln(1+s/b)-s]}, \\
 \mathcal{Z}_\text{excl} &= \sqrt{2[s-b\ln(1+s/b)]}.
\ee
Here $s$ and $b$ denote the event numbers after the above cuts for the signal and background, respectively. $\delta$ denotes the percentage systematic error on the SM
background estimate. The integrated luminosity
at the HL-LHC is set at 3000~fb$^{-1}$.

%%% Fig.5 %%%%%%%%%%%%%%%%%%%%
\begin{figure}[htb]
\begin{center}
\vspace{1.5cm}
\centerline{\epsfxsize=7cm \epsffile{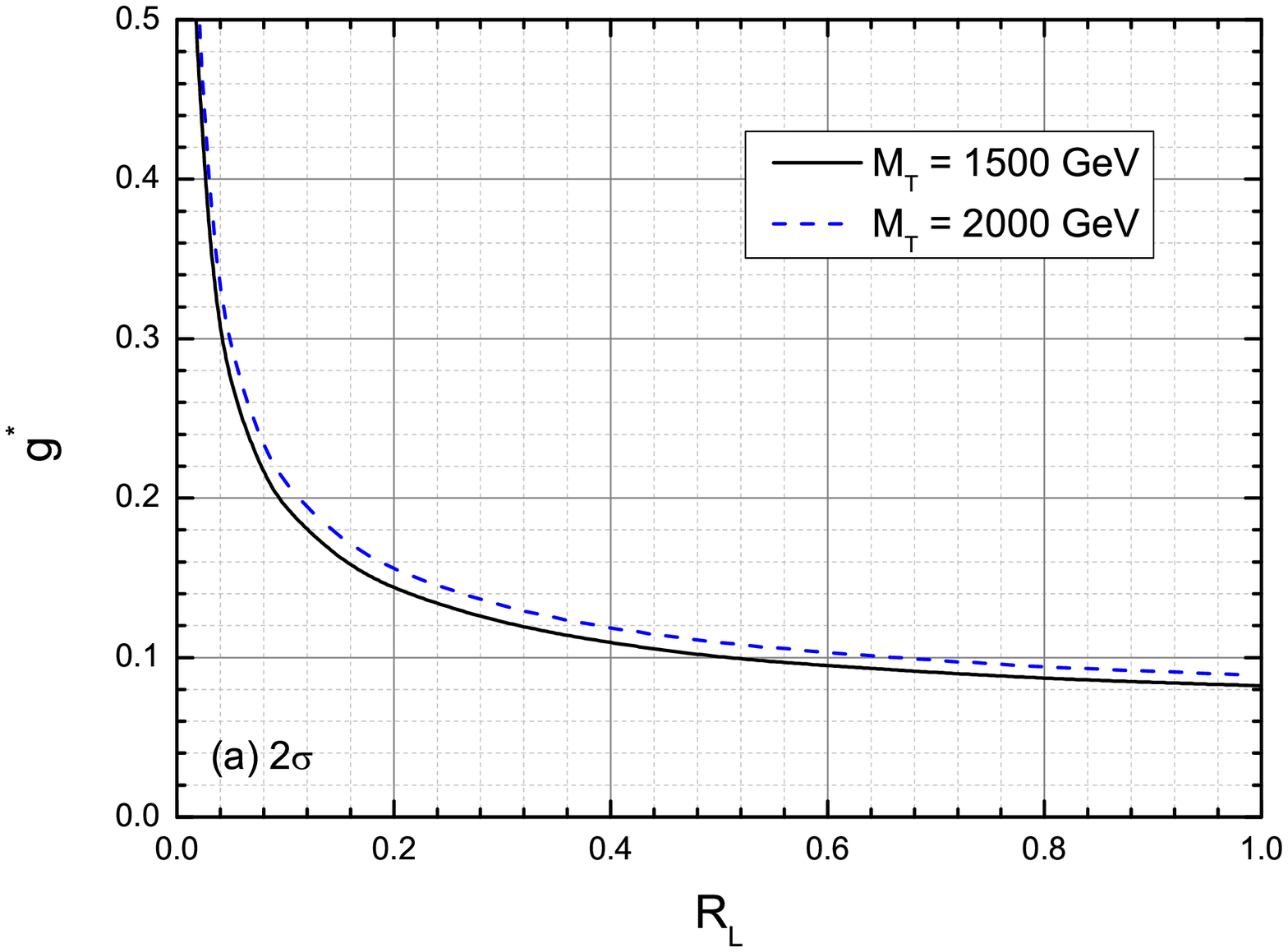}\epsfxsize=7cm \epsffile{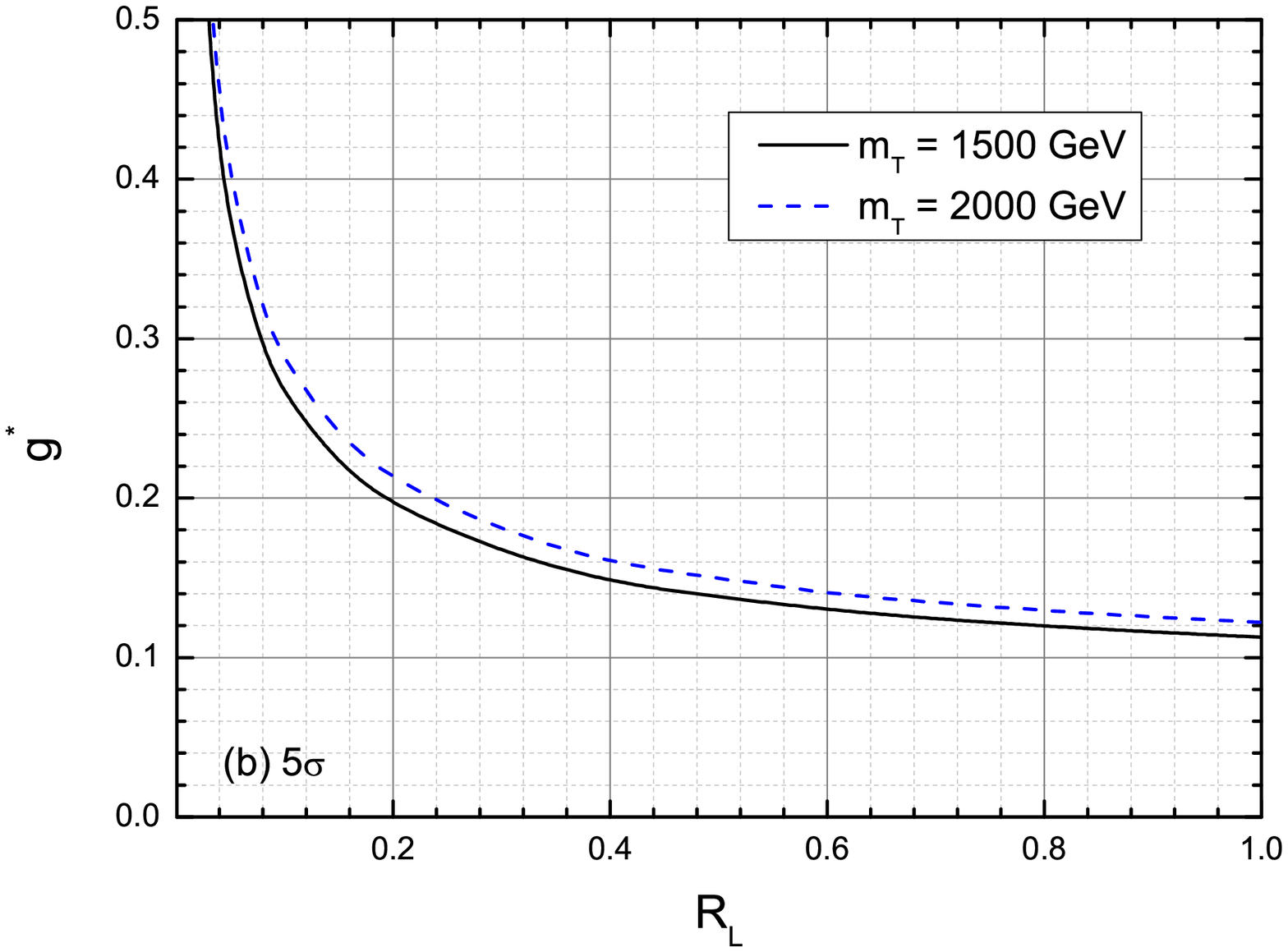}}
\caption{$2\sigma$ (left panel) and $5\sigma$ (right panel) contour plots for the signal in $g^{\ast}-R_L$ with two typical VLQ-$T$ masses at HL-LHC.  Here we consider a systematic
uncertainty of $\delta=30\%$.  }
\label{ss}
\end{center}
\end{figure}

In Fig.~\ref{ss}, we plot the excluded $2\sigma$ and $5\sigma$ discovery reaches in the plane of $g^{\ast}-R_L$ for two  fixed VLQ-$T$ masses and $\delta=30\%$ at HL-LHC. In Fig.~\ref{ss}, one can see that the $5\sigma$ level discovery sensitivity of $g^{\ast}$ is 0.11~(0.12) for $M_T=1500~(2000)$ GeV and $R_L=1$, and it changes as $0.27~(0.29)$ for $R_L=0.1$. On the other hand, from the $2\sigma$ exclusion limits one can see that the upper limits on the size of $g^{\ast}$ are given as $g^{\ast}\leq 0.08~(0.09)$ for $R_L=1$, and that they change as $g^{\ast}\leq 0.42~(0.45)$ for the smaller value $R_L=0.02$.

 %%%%%%%%%%%%%%%%%%%%%%%
\begin{figure}[!htb]
\begin{center}
\centerline{\epsfxsize=10cm \epsffile{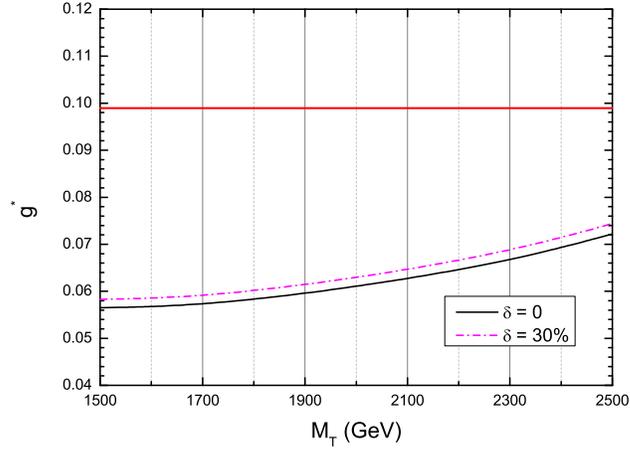}}
\caption{$2\sigma$ contour plots for the signal in $g^{\ast}-M_T$ planes at HL-LHC with different values of the systematic uncertainty, assuming that the VLQ-$T$ couples only to first-generation SM quarks.  The lower bounds from non-LHC flavor physics are indicated by the red horizontal contour.}
\label{fig6}
\end{center}
\end{figure}

 As mentioned earlier,  the case of $R_L\rightarrow\infty$  means that the singlet VLQ-$T$ is coupled only to the first-generation SM quarks. Based on the cuts adopted in the above discussion, we extend our analysis in this case with the VLQ-$T$ masses ranging from 1500 to 2500~GeV in steps of 100~GeV. Figure~\ref{fig6} shows the $2\sigma$ exclusion limits as a function of $M_T$ and
$g^{\ast}$  with two systematic error cases of  $\delta=0$ and $\delta=30\%$. We observe that  our signals are not very sensitive to the values of the systematic uncertainties.
 Assuming a realistic 30\% systematic error, the sensitivities are slightly weaker than those without any systematic error. For the considered mass range of 1500 to 2500~GeV, the upper limit on allowed values of $g^{\ast}$ rises from a
minimum value of 0.056 starting at $M_T=1500$~GeV, up to 0.074 for $M_T=2500$~GeV. These results are slightly better than the noncollider limits~($\kappa=g^{\ast}/\sqrt{2}\simeq 0.07$) conservatively estimated in Ref.~\cite{Buchkremer:2013bha} for a mass scale
of the order of a TeV from atomic parity violation measurements~\cite{Deandrea:1997wk}.

\section{CONCLUSION}
The new heavy vectorlike $T$ quark of charge 2/3 appears in many new physics models beyond the SM.
In this paper, we exploited a simplified model with only two free parameters: the electroweak coupling parameter $g^{\ast}$ and the generation mixing parameter $R_L$. We presented a search strategy at the future HL-LHC for a distinguishable signal with a same-sign dilepton plus two jets and missing energy.  The $2\sigma$ exclusion limits, as well as the $5\sigma$ discovery reach in the parameter plane of the two variables  $g^{\ast}-R_L$, were obtained for two typical heavy $T$ quark masses. For two typical VLQ-$T$ masses $M_T=1500~(2000)$ GeV, the upper limits on the size of $g^{\ast}$ were given as $g^{\ast}\leq 0.42~(0.45)$ for the smaller value $R_L=0.02$, and $g^{\ast}\leq 0.08~(0.09)$ for $R_L=1$. Assuming that the VLQ-$T$ with mass of TeV scale couples to the first-generation quarks only,  the correlated region $g^{\ast}\in[0.056,0.074]$ and $M_T\in [1500,2500]$~GeV can be excluded at the $2\sigma$ level at the future HL-LHC, which is slightly better than the noncollider limits from atomic parity violation measurements.

\begin{acknowledgments}
This work is supported by the key research and development program of Henan Province~(Grant No.~22A140019) and the Natural Science Foundation of Henan Province~(Grant No.~222300420443).
\end{acknowledgments}

%%%%%%%%%%%%%%%%%%%%%%%%%%%%%%%%%%%%%%%%%%%%%%%%%%%%%%%%%%%%%%%%%%%%%%%%%%%%%%%%%%
%                                        Appendix
%%%%%%%%%%%%%%%%%%%%%%%%%%%%%%%%%%%%%%%%%%%%%%%%%%%%%%%%%%%%%%%%%%%%%%%%%%%%%%%%5

\end{document}